\documentclass[preprint2]{aastex}

\begin{document}

\title{The distance and radius of the neutron star PSR B0656+14}

\author{Walter F. Brisken,\altaffilmark{1} 
	S. E. Thorsett,\altaffilmark{2} 
	A. Golden,\altaffilmark{3} \&
	W. M. Goss\altaffilmark{1}} 
\altaffiltext{1}{National Radio Astronomy Observatory, PO Box O, Socorro, 
NM 87801; \email{wbrisken@nrao.edu}} 
\altaffiltext{2}{Department of Astronomy and Astrophysics, University of
		California, Santa Cruz, CA 95064}
\altaffiltext{3}{National University of Ireland, Newcastle Road,
Galway, Republic of Ireland}

\begin{abstract}
We present the result of astrometric observations of the radio pulsar
PSR B0656+14, made using the Very Long Baseline Array.  The parallax
of the pulsar is $\pi=3.47\pm0.36$~mas, yielding a distance
$288^{+33}_{-27}$~pc. This independent distance estimate has been used
to constrain existing models of thermal x-ray emission from the
neutron star's photosphere. Simple blackbody fits to the x-ray data
formally yield a neutron star radius $R_\infty\sim7-8.5$~km. With more
realistic fits to a magnetized hydrogen atmosphere, any radius between
$\sim13$ and $\sim20$~km is allowed.
\end{abstract}


\section{Introduction}

The cooling of neutron stars offers a unique diagnostic of the physics
of the interior. Because the emergent spectrum is nearly blackbody,
the observed flux can be modeled to estimate the temperature,
the intervening column of absorbing material, and the ratio
of radius to distance. 
Astrometric measurement of the distance then allows an
estimation of the photospheric radius, which is sensitively
dependent on the high density equation of state.

PSR~B0656+14 is a middle-aged ($\sim 10^5$~yr old) pulsar, with a
spectrum dominated at ultraviolet and soft-X-ray wavelengths by a
$\sim10^6$~K blackbody component, as expected in standard cooling
models. Distance estimates obtained from measurements of radio
dispersion in the ionized material between Earth and the pulsar have
led to a large estimated stellar radius, $>20$~km.  Here we show that
the parallax of PSR B0656+14, determined using very long baseline
interferometry, leads to a factor three reduction in the estimated
pulsar distance. We place this result in the context of existing
thermal models for the neutron star.

\section{VLBA Observations \& Astrometric Analysis}

For many years, pulsar parallax measurements have been just at or
beyond experimental limits. The situation has greatly improved with
the increased sensitivity available using the NRAO\footnote{The
National Radio Astronomy Observatory is a facility of the National
Science Foundation operated under cooperative agreement by Associated
Universities, Inc.} Very Long Baseline Array (VLBA), together with new
phase-referencing techniques for measuring and eliminating ionospheric
distortion and the ability to gate the VLBA correlator to accept data
only during intervals when the pulsar is ``on.'' In the last three
years, parallaxes (and distances) to ten pulsars have been measured
with the VLBA (Brisken et~al.\ 2002; Chatterjee et~al.\ 2001).

PSR B0656+14 has a 1.4~GHz flux density of 3.6~mJy, which is too weak
to use the ionosphere calibration technique described by Brisken et
al.\ (2002).  Instead, the in-beam calibration technique used by
Chatterjee et al.\ (2001) was applied.  Very Large Array (VLA)
observations were made in the A-configuration at 1.4~GHz to search for
potential in-beam calibrators.  Three candidates were followed up at
8.4 and 15~GHz to test for compactness.  Fortunately one of them,
0658+1410, was compact and bright enough for use as an in-beam
calibrator.  See Table~\ref{tab:calibrators} for a list of calibrators
used, their coordinates, and their flux densities.

Five observations of PSR B0656+14, each five hours long, were made over
1.3~yrs with the VLBA.  A 32~MHz band centered at 1.6675~GHz was used,
resulting in a beam size of $15 \times 6$~mas.  Each observation
consisted of sixty 2~minute scans on the field containing the pulsar
and in-beam calibrator, each bracketted by observations of J0637+1458,
and two 5~minute scans on 3C147 for bandpass calibration.  The target
field was correlated twice.  The first correlation placed the in-beam
calibrator at the phase center.  The second put the pulsar at the
phase center and used the pulsar gate to improve the
signal-to-noise.

Considerable effort went into ensuring that each observation was
reduced identically. Analysis used the standard AIPS package.  After
editing the data to remove radio-frequency interference, data from all
five epochs were combined to make a single model for each calibrator
source using standard techniques.  After initial amplitude and
bandpass calibration, GPS-based ionosphere calibrations were applied
with the task TECOR.  J0637+1458 was then fringe fitted (with the
model) to determine the delays and rates and to perform a first-order
phase calibration, which was followed by phase-and-amplitude
self-calibration.  Calibration derived from J0637+1458 was
interpolated in time and applied to 0658+1410.  This source was
self-calibrated for phases only, again with a model.  The calibrations
from both J0637+1458 and 0658+1410 were applied to the pulsar, images
were produced, and image-plane Gaussian fitting performed to determine
the pulsar position.

Uncertainties are dominated by systematic effects, mainly due to
atmospheric and ionospheric refraction.  To estimate the size of these
effects, imaging was repeated at each epoch with subsets of the data.
The measured position varied most with varying time cuts. The scatter
in position when varying the time ranges was used to assign error
bars.  Final single-epoch position uncertainties were 0.8~mas and
0.4~mas in right ascension and declination respectively.  The position
of B0656+14 at each epoch is shown in Table~\ref{tab:positions}.  Note
that the absolute positions are accurate only to about 10~mas due to
the uncertainty in the true position of the calibrator 0658+1410.

\begin{figure}
\plotone{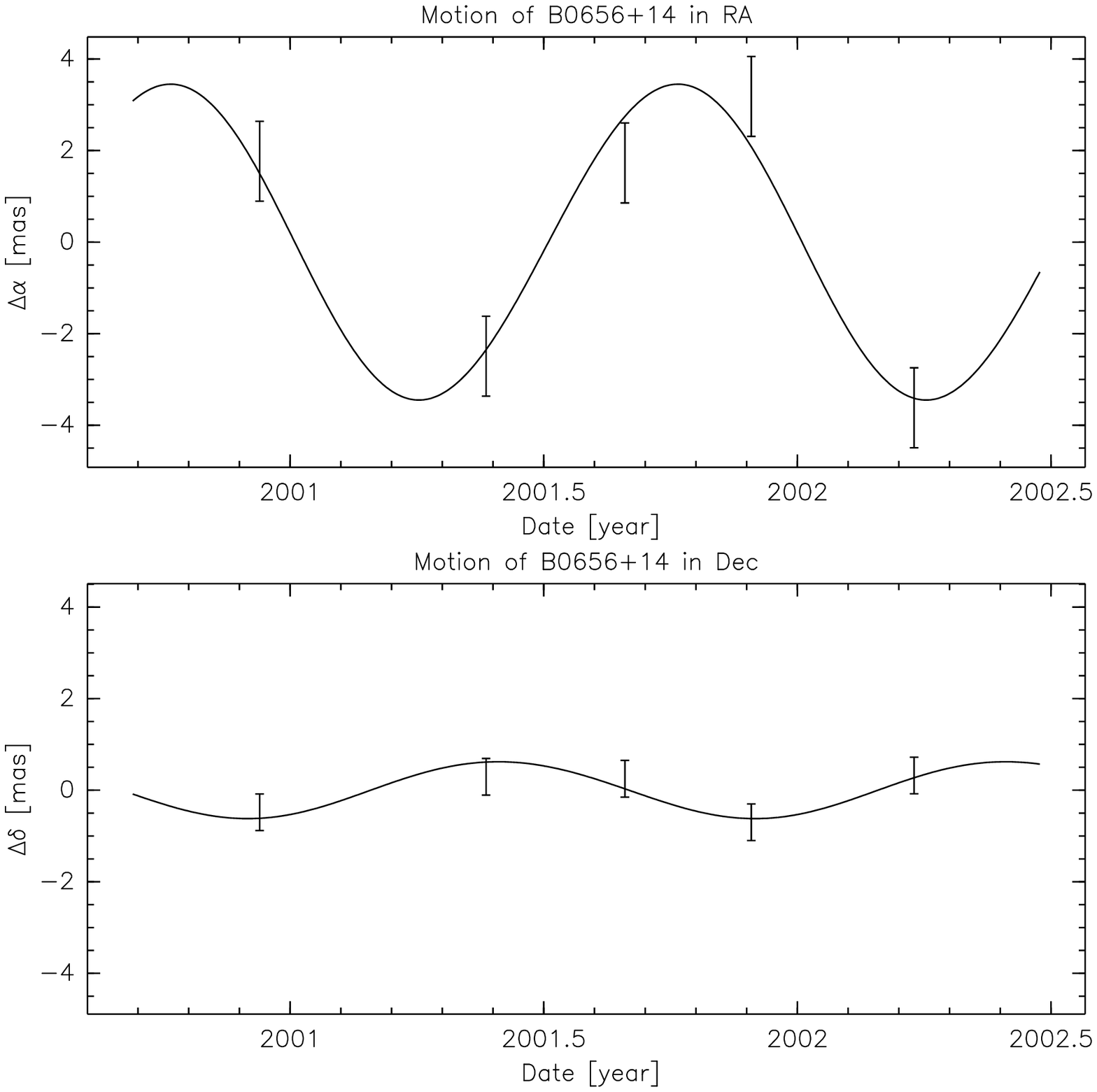}
\caption[f1.eps]{\label{fig:pi}
The parallax fit to five measured positions of B0656+14.  The proper motion
was removed from the data and model.
}
\end{figure}

The five measured positions were used to simultaneously fit the
parallax, $\pi$, proper motion ($\mu_{\alpha} \cos(\delta)$ and
$\mu_{\delta}$), and position of PSR B0656+14.  These values and the
derived distance, transverse velocity, $v_{\perp}$, and line-of-sight
column density of electrons, $n_{\mathrm{e}}$ are shown in
Table~\ref{tab:derived}.  Figure~\ref{fig:pi} shows the parallax fit
to the data points.  In this plot, the proper motion has been removed
from the data to allow the subtle parallax to be seen.

The proper motion measured here differs from that measured by Thompson
and C\'ordova (1994), $\mu_\alpha\cos\delta=64\pm11$ and
$\mu_\delta=-28\pm4$~mas/yr, by more than $6\sigma$. It is in
excellent agreement with the {\it HST}-derived value,
$\mu_\alpha\cos\delta=42.7\pm2$ and $\mu_\delta=-2.1\pm3$~mas/yr
(Mignani, De Luca, \& Caraveo 2000), though somewhat more precise.

\section{Discussion}

\subsection{Implications for Galactic electron density model}

The distance determined from the pulsar dispersion measure (DM) using
the Galactic electron distribution model of Taylor and Cordes (1993)
is $760\pm190$~pc. Using the electron distribution model of Cordes and
Lazio (2002) yields $669\pm73$~pc. Both values are substantially
greater than the parallax distance, $288^{+33}_{-27}$~pc. The
discrepancy is not entirely surprising---the galactic electron density
is poorly sampled at this $(l,b)$ due to a lack of calibration sources
near the $201^{\circ}$ galactic latitude of PSR~B0656+14 (Anderson et
al. 1993).  Furthermore, the nearest calibration source for the Taylor
\& Cordes model, PSR~B0823+26 at $l = 197^{\circ}$, is known to have a
DM derived distance some 3 times its measured parallax (Gwinn et
al. 1986). Nevertheless, the high mean electron density along this
path, $n_e=0.049$~cm$^{-3}$, is a reminder that dispersion
distances are particularly uncertain in the Solar neighborhood, where
lines of sight do not average over many independent regions.

\subsection{Implications for thermal modeling and the stellar radius}

PSR~B0656+14 has been extensively studied at optical (Shearer et al.\
1997; Kurt et al.\ 1998; Koptsevich et al.\ 2001), ultraviolet
(Pavlov, Stringfellow, \& C\'ordova 1996; Pavlov, Welty, \& C\'ordova
1997; Edelstein et al.\ 2000), x-ray (C\'ordova et al.\ 1989; Finley,
\"Ogelman, \& Kizilo\u{g}lu 1992; Anderson et al.\ 1993; Greiveldinger
et al.\ 1996; Mineo et al.\ 2002; Marshall \& Schulz 2002; Pavlov et
al.\ 2002), and gamma-ray (Ramanamurthy et al. 1996) energies. 

A simple model that adequately accounts for observations from the
optical through the x-ray bands includes a power law component,
identified with non-thermal emission from the active magnetosphere,
and two blackbody components: a harder component identified with the
heated polar caps, and a softer component identified with emission
from the entire photosphere. In general, model fits agree very well.
For example, a model based on the {\it EUVE, ROSAT,} and {\it ASCA}
data (Koptsevich et al.\ 2001) gives $T_\infty=8.4\pm0.3\times10^5$~K
for the temperature of the soft component, while an independent model
based on {\it Chandra} observations (Marshall \& Schulz 2002) gives
$T_\infty=8.0\pm0.3\times10^5$~K, where $T_\infty$ is the gravitationally
redshifted temperature measured by a distant observer:
$T_\infty=T(1-2GM/Rc^2)^{1/2}$.  For our distance of 288~pc, the
resulting estimates for the radius of the neutron star,
$R_\infty=R(1-2GM/Rc^2)^{-1/2}$ are 7.8 and 8.5~km,
respectively. Including the optical/UV data and forcing a single
power-law slope for the non-thermal component across more than four
orders of magnitude in energy, Pavlov et al.\ (2002) find a slightly
smaller radius, 6.9~km at 288~pc.  Estimates for the radius of the hot
polar cap vary from 0.5 to 0.64~km at that distance.

A blackbody radius as small as $R_\infty=8$~km would pose serious
challenges to standard neutron star models (e.g., Heiselberg \&
Pandharipande 2000), though it would be acceptable for a low mass
strange star (Alcock, Fahri, \& Olinto 1986).  However, a more likely
explanation for this anomalous result is failure of the simple
black-body modeling.\footnote{There has also been some concern that
flux from an apparent pulsar wind nebula that has been barely resolved
by {\it Chandra} might contaminate the x-ray spectra (Pavlov et al.\
2002). Because the estimated luminosity is less than 4\% of the
bolometric luminosity of the soft thermal component, any systematic
effect on the radius estimation is likely to be small, but larger
errors might be introduced to the polar cap radius.}

In general, realistic atmosphere models tend to shift flux into high
energy tails that result in overestimation of the effective
temperature and hence underestimation of the stellar radius (see
Romani 1987 and Shibanov et al.\ 1992 for early work on non-magnetic
and magnetic atmospheres, respectively, and Zavlin \& Pavlov 2002 for
a recent review). The largest radii are found with non-magnetic
hydrogen or helium atmospheres, while strong magnetic fields suppress
the high energy tails and produce intermediate temperature and radius
estimates.  Our blackbody model radius should therefore most properly
be treated as a lower limit to the neutron star radius.

For PSR~B0656+14, fits to unmagnetized hydrogen atmospheres (e.g.,
Marshall \& Schulz 2002) produce very large radius estimates: $\sim
156$~km at our measured distance. This can be regarded as an
(unconstraining) upper limit to the radius.

Magnetized atmospheres produce much more reasonable results. Using
just the {\it ROSAT} data, the magnetic hydrogen models of Shibunov et
al. (1993), and a canonical neutron star radius of $R_\infty=13$~km,
Anderson et al.\ (1993) estimated the pulsar distance to be
$280^{+60}_{-50}$~pc, in remarkable agreement with our
measurement. Scaling from model fits done for the Geminga pulsar,
Marshall \& Schulz (2002) estimate that the using a magnetized
hydrogen atmosphere will increase the radius by a factor 2.3 over a
blackbody estimate, resulting in corrected radius estimates of 15.9,
17.9, and 19.7~km for the Pavlov et al.\ (2002), Koptsevich et al.\
(2001), and Marshall \& Schulz (2002) model fits, respectively.  We
conclude that within the model uncertainties, any radius $R_\infty$
between $\sim13$ and $\sim20$~km is allowed.

\subsection{The next steps}

A promising observational approach is to complement the high-energy
observations with improved constraints on the flux in the optical and
ultraviolet bands.  Anderson et al.\ (1993) found that the
blackbody and magnetized hydrogen atmosphere models that each
fit the {\it ROSAT} data have significantly different predictions in
the optical bands: for example, $B\approx28.7$ and $B\approx27.8$,
respectively. 

Analysis of the combined integrated photometric
data indicate that the spectral energy distribution is predominantly
thermal in the ultraviolet, then flattens out and asymptotically approaches a
power-law-like form towards the infrared (Koptsevich et al. 2001).  The
photometry is well fit by a Rayleigh-Jeans plus synchrotron-like
power-law model---the former agreeing to first order with the X-ray
extrapolations. The power-law, with a best-fit index $\sim1.5$, is
similar to that observed at hard-x-ray and gamma-ray energies.

In the ultraviolet, comparisons with model predictions are challenging because
of the uncertain reddening.  In the optical, the challenge has been to
detect the thermal contribution beneath the larger non-thermal
power-law. The observed $B$ magnitude is $24.85^{+0.19}_{-0.16}$ (Kurt
et al. 1998), more than an order of magnitude above the expected
thermal contribution.

Of considerable interest, then, has been the detection of optical
pulsations.  Shearer et al. (1997) reported pulsations based on
$B$-band observations with a 2-d counting MAMA camera, and Kern (2002)
detected pulsations in the $\sim V$ band using a phase-binning CCD
system.  In addition to confirming the magnetospheric origin of the
optical radiation, these observations raise the possibility of
measuring and removing the non-thermal component.  While there are
clear differences in the light curve morphologies, both datasets are
consistent with the radiation being 100\% pulsed. At this time, only
upper limits can be placed on any unpulsed thermal emission: in $B$,
the limit is $B\leq26.8$.  This is within a factor $\sim3$ of the
predicted value.  Of considerable interest would be extending these
observations to $U$ band, where the nonthermal flux will be smaller
and the predicted thermal flux is a factor four higher.

These observations will be challenging, as will continued improvement
in the magnetic atmosphere modeling.  But both observations and theory
have seen remarkable improvement over the last decade, even when the
ultimate limit was the very uncertain pulsar distance. Now the
distance to PSR B0656+14 is known to 10\%, comparable to the typical
uncertainty of individual spectral model fits, and somewhat below the
systematic uncertainties in the atmospheric fitting. There is no
reason that continued radio observations shouldn't continue to improve
the parallax limit: precision a factor five better has now been
achieved for PSR B0950+08.  Thus there is every reason to believe that
PSR B0656+14 will continue to develop into a fascinating astrophysical
laboratory for nuclear physics.

\acknowledgements

We wish to thank Andrew Lyne for providing us with the pulsar ephemeris
that allowed us to gate the VLBA correlator. S.E.T. is supported by
the NSF under grant AST-0098343.  A.G. is supported by Enterprise Ireland 
under grant SC/2001/322.



\begin{deluxetable}{lllrrl}
\tablewidth{0pt}
\tablecaption{Calibrator sources\label{tab:calibrators}}
\tablehead{ 
    \colhead{Object}
  & \colhead{$\alpha_{\mathrm{J2000}}$}
  & \colhead{$\delta_{\mathrm{J2000}}$}
  & \colhead{Peak ${S_{1667}}^{\dagger}$}
  & \colhead{Int. ${S_{1667}}^{\dagger}$}
  & \colhead{Remarks}
 \\ 
  & 
  & 
  & \colhead{(mJy/beam)}
  & \colhead{(mJy)}
  &
}
\startdata
3C147      & $05^h42^m36^s.1379$ & $49^{\circ}51'07''.234$ & 1493 & 2998 & 
		Bandpass calibrator \\
J0637+1458 & $06^h37^m51^s.0523$   & $14^{\circ}58'57''.278$   & 234  & 250 & 
		Phase calibrator \\
0658+1410  & $06^h58^m46^s.285$    & $14^{\circ}10'37''.83$  & 33   & 38 & 
		In-beam calibrator \\
\enddata
\tablenotetext{\dagger}{Peak and integrated flux densities are averaged over
the 1651.5 to 1683.5~MHz observed band.}
\end{deluxetable}

\begin{deluxetable}{cll}
\tablewidth{0pt}
\tablecaption{B0656+14 position measurements\label{tab:positions}}
\tablehead{
    \colhead{Date} 
  & \colhead{$\alpha_{\mathrm{J2000}}$}
  & \colhead{$\delta_{\mathrm{J2000}}$}
}
\startdata
2000.940 & $06^h59^m48^s.15018$ &  $14^{\circ}14'21''.1576$ \\ 
2001.386 & $06^h59^m48^s.15124$ &  $14^{\circ}14'21''.1573$ \\ 
2001.660 & $06^h59^m48^s.15236$ &  $14^{\circ}14'21''.1566$ \\ 
2001.909 & $06^h59^m48^s.15322$ &  $14^{\circ}14'21''.1551$ \\ 
2002.230 & $06^h59^m48^s.15372$ &  $14^{\circ}14'21''.1553$ \\ 
\enddata
\end{deluxetable}

\begin{deluxetable}{ll}
\tablewidth{0pt}
\tablecaption{Derived parameters for B0656+14\label{tab:derived}}
\tablehead{
    \colhead{Parameter}
  & \colhead{Value$^{\dagger}$}
}
\startdata
${\alpha_{\mathrm{J2000}}}^{\ddag}$ & $06^h59^m48^s.1472$ \\
${\delta_{\mathrm{J2000}}}^{\ddag}$ & $14^{\circ}14'21''.160$ \\
$\mu_{\alpha} \cos(\delta)$ & $44.07 \pm 0.63$ mas yr$^{-1}$ \\
$\mu_{\delta}$ & $-2.40 \pm 0.29$ mas yr$^{-1}$ \\
$\pi$ & $3.47 \pm 0.36$ mas \\
Distance & $288^{+33}_{-27}$ pc \\
$v_{\perp}$ & $60^{+7}_{-6}$ km s$^{-1}$ \\
$n_{\mathrm{e}}$ & $0.048 \pm 0.005$ cm$^{-3}$ \\
\enddata
\tablenotetext{\dagger}{Uncertainties are 68\% confidence intervals.}
\tablenotetext{\ddag}{The pulsar position is relative to the frame 
defined by 0658+1410.  The uncertainty in the coordinates of 0658+1410
dominate the total uncertainty of the pulsar's absolute position and 
are estimated to be about 10~mas.  }
\end{deluxetable}

\end{document}